\documentclass[twocolumn,showpacs,preprintnumbers,amsmath,amssymb]{revtex4}
\usepackage{graphicx}
\usepackage{dcolumn}
\usepackage{bm}

\begin{document}


\title{A New Equation for Describing Heat Conduction by Phonons}

\author{K. Razi Naqvi and S. Waldenstr\o m}
\affiliation{Department of Physics,
Norwegian University of Science and Technology,
NO-7491 Trondheim, Norway}

\date{\today}

\begin{abstract}
A new equation, rooted in the theory of Brownian motion, is proposed for describing heat conduction by phonons. Though a finite speed of propagation is a built-in feature of the equation, it does not give rise to an inauthentic wave front that results from the application of the hyperbolic heat equation (of Cattaneo). Even a simplified, analytically tractable version of the equation yields results close to those found by solving, through more elaborate means, the equation of phonon radiative transfer. An explanation is given as to why both Brownian motion and its inverse (radiative transfer) provide equally serviceable paradigms for phonon-mediated heat conduction.  

\end{abstract}

\pacs{44.10.+1, 05.40.-a, 05.60.Cd, 66.70.+f}
\maketitle

Thermal conduction has been treated, until quite recently, within the framework of the classical parabolic heat equation, associated with the name of Fourier, or by using the hyperbolic heat equation, frequently named after Cattaneo \cite{Luikov:68,JoPrez:RMP99}. The failure of Fourier's equation, which becomes apparent when one examines heat conduction on small scales \cite{Maju:93,JoshMaju:93,ChenPRL:01,ChenASME:02}, is easily grasped, for it implies an infinite speed of heat propagation; more disappointing, however, is the failure, under similar circumstances, of Cattaneo's equation, since it does assign a finite value to the speed of propagation. Recognition of the inadequacies of these equations prompted Majumdar \cite{Maju:93} to introduce an equation that has come to be known as the equation of phonon radiative transfer (EPRT); it models conduction of heat in thin dielectric films as the transport of phonons, in analogy with the linear Boltzmann equation (LBE) used for describing the transport of photons and monoenergetic neutrons \cite{DudeMart:79}. Since even simple versions of EPRT are time-consuming \cite{JoshMaju:93}, one is tempted to replace it with an easier alternative that is more trustworthy than the equations of Fourier and Cattaneo. Recently, Chen has proposed an approach where Cattaneo's equation is grafted onto the collisionless form of EPRT \cite{ChenPRL:01, ChenASME:02}. Chen's recipe, to be labeled here as the ballistic-diffusive approximation (BDA), appears to combine simplicity with reliability, and his results are highly encouraging. An unsatisfactory feature of BDA, traceable to Chen's employment of Cattaneo's equation, is the occurrence of an artificial wave front in the diffusive component of the internal energy (and therefore in the total internal energy as well). Another manifestation of the distortion introduced by Cattaneo's equation is a small but significant disagreement between the temperature profiles predicted by EPRT and BDA. 

In this Letter we propose a new equation for describing phonon-mediated heat conduction, and present results obtained by employing a simpler, analytically tractable version of the equation. Since we wish to compare our results with those reported by other workers \cite{JoshMaju:93, ChenPRL:01, ChenASME:02}, we will treat a plane-parallel conducting medium. We will denote the average speed of sound by $\overline{v}$, and the mean free path of phonons by $\overline{l}$; the symbol $\overline{\tau}\equiv \overline{l}/\overline{v}$ will be called the relaxation time and $1/\overline{\tau}$ will be denoted by $\alpha$.  

The complete new equation will be presented after we have considered a simplified, exactly soluble variant, and examined its performance; meanwhile we will use the label NHE for the simpler version, which can be written, with $\partial_y\equiv \partial/\partial y$ and $\partial_{yy}\equiv \partial_y\partial_y$, as
\begin{equation}
\partial_t T(x,t)=(1-e^{-\alpha t})\kappa\, \partial_{xx} T(x,t),
\end{equation}
where $T(x,t)$ denotes the temperature at time $t$ at the point $x$, and $\kappa= \overline{v}\overline{l}/3$ is thermal diffusivity. One sees immediately that Eq.~(1) reduces to PHE, $\partial_t T=\kappa \partial_{xx}T$, in the long-time limit ($e^{-\alpha t}\ll 1$). We also point out that the substitution
\begin{equation}
s=t-{1-e^{-\alpha t}\over \alpha}
\end{equation}
reduces Eq.~(1) to $\partial_s T=\kappa \partial_{xx} T$. 

To begin with the statement of the problem: A slab of thickness $L$ is initially at a uniform temperature $T_0$; at time $t=0$, one face (say that at $x=0$) is raised to a temperature $T_1$ and is maintained at this temperature thereafter, the other face (at $x=L$) being kept at the temperature $T_0$; we wish to find the temperature $T(x,t)$ and the flux $q(x,t)$, for $t>0$ and $0\leq x \leq L$. Since the problem has already been treated in the past \cite{JoshMaju:93, ChenPRL:01, ChenASME:02}, we will not give more details other than spelling out our notation: $\Delta T=T_1-T_0$, $\xi=x/L$, $t^\ast=\alpha t$, $s^\ast=\alpha s$, $\theta =(T-T_0)/\Delta T$, $\phi=q/(Cv\Delta T)$, ${\rm Kn}=\overline{l}/L$; $C$ stands for the specific heat per unit volume. 

It will be convenient to append to $\theta$ and $\phi$ an appropriate suffix (P, H, or N, depending on whether the result pertains to the parabolic, hyperbolic, or new heat equation). The equations used for calculating $\theta_P(\xi,t^\ast)$, $\theta_H(\xi,t^\ast)$, $\phi_P(\xi,t^\ast)$, $\phi_H(\xi,t^\ast)$ have been stated (with some misprints) by previous authors \cite{JoshMaju:93, ChenASME:02}. The transformation shown in Eq.~(2) enables us to express $\theta_N(\xi,t^\ast)$ and $\phi_N(\xi,t^\ast)$ as follows:
\begin{eqnarray}
\theta_N(\xi,t^\ast)&=&\theta_P(\xi,s^\ast),\\
\phi_N(\xi,t^\ast)&=&(1-e^{- t^\ast})\phi_P(\xi,s^\ast).
\end{eqnarray}

Figure 1 compares the performance of NHE with the classical equations; here the nondimensional temperature $\theta (\xi,t^\ast)$ and the nondimensional heat flux $\phi (\xi,t^\ast)$  are plotted against $\xi$ at $t^\ast =1$ in a slab for which ${\rm Kn}=1$. A similar comparison (involving Fourier's equation, Cattaneo's equaton, BDA and EPRT) has been made by Chen \cite{ChenPRL:01, ChenASME:02}; the coordinates of his EPRT plots have been read off and included in Fig.~1 to facilitate comparison. Chen has rescaled his data so as to harmonize the definitions pertaining to the different equations; in the absence of such rescaling the EPRT plots show jumps at the boundaries (see below). One sees that the temperature profile predicted by NHE is in excellent agreement with the rescaled EPRT plots; the flux predicted by NHE deviates visibly from the EPRT plot (for $\xi > 0.7$), but the other two plots (PHE and HHE) are in severe disagreement with the EPRT plot for nearly all values of $\xi$. 

\begin{figure}
\vspace*{-7.5 truemm}
\hspace*{-7.5 truemm}
\includegraphics[height=13cm,width=10cm,angle=0]{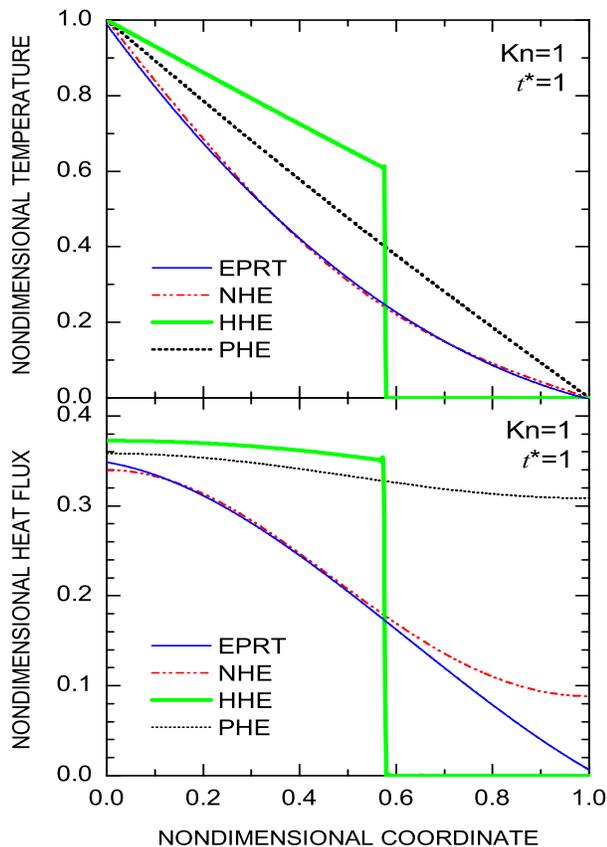}
\vspace*{-10 truemm}
\caption{\label{fig:Figure1} Comparison of the spatial profiles of $\theta(\xi,t^\ast)$ and $\phi(\xi,t^\ast)$ calculated by using the equation of phonon radiative transfer (EPRT), the parabolic heat equation of Fourier (PHE), the hyperbolic heat equation of Cattaneo (HHE) and Eq.~(1), the new heat equation (NHE). The data for the curve labeled EPRT have been taken from Refs. \cite{ChenPRL:01, ChenASME:02}.}
\end{figure}

Curves showing the surface heat flux $\phi (0,t)$ are plotted in Fig.~2, and compared with reconstructions of Chen's plots. The top panel refers to a slab that is so thin that transport becomes essentially ballistic---a situation that can be better handled by using the extended version of Eq.~(1) to be presented shortly. In the middle and bottom panels of Fig.~2, one notes a persistent difference between NHE and EPRT; this is to be expected because the EPRT data originate from plots which have not been rescaled.

\begin{figure}
\vspace*{-7.5 truemm}
\includegraphics[height=16cm,width=7.2cm,angle=0]{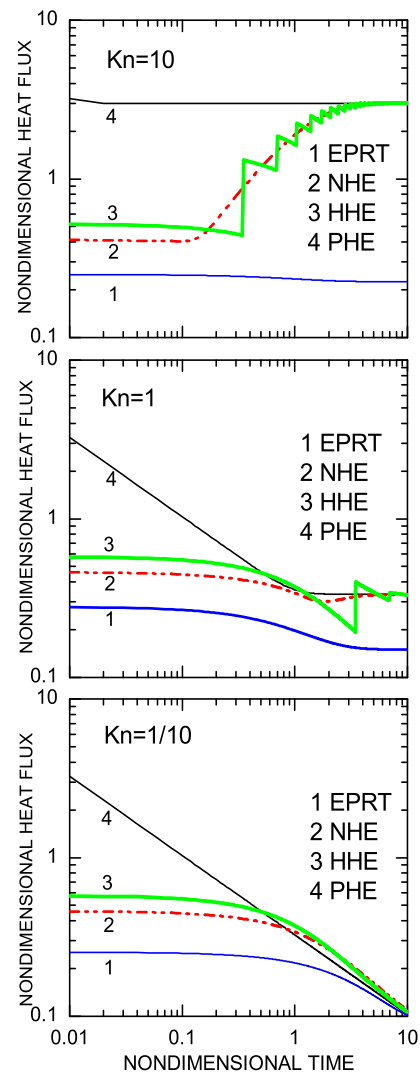}
\vspace*{-10 truemm}
\caption{\label{fig:Figure2} Comparison of the temporal profile of the heat flux at $\xi=0$ calculated by using EPRT, the PHE, HHE and NHE; for an explanation of the labels refer to Fig.~1. The EPRT data have been read from Refs. \cite{ChenPRL:01, ChenASME:02}.}
\end{figure}

Let us pause briefly to note that NHE provides, in all cases, a better description than that furnished by Cattaneo's equation. This means that the performance of BDA can be improved by discarding the Cattaneo equation in favor of NHE. We return now to NHE and investigate the prospects of using it as a complete equation in itself; for this purpose, we will not consider samples where transport is dominated by ballistic behavior.

One would expect, in view of some previous investigations \cite{Maju:93, JoshMaju:93, Klis:88}, temperature jumps to occur at the boundaries of sufficiently thin samples. To cope with temperature jumps, we must replace the boundary conditions $\theta(0,t^\ast)=1$ and $\theta(1,t^\ast)=0$ with $\theta(-\xi_0,t^\ast)=1$ and $\theta(1+\xi_1,t^\ast)=0$, respectively.  Postponing the issue of how the values of the {\it extrapolated endpoints\/}  ($x_0\equiv L\xi_0$ and $x_1\equiv L\xi_1$) are to be found, we will estimate their values by using the results published by Chen \cite{ChenASME:02}; it is pertinent to recall that, in the steady state, the diffusive part of his solution satisfies Marshak's boundary condition \cite{DudeMart:79}: $x_0=x_1=2\overline{l}/3$, or equivalently $\xi_0=\xi_1=2{\rm Kn}/3$.

\begin{figure}
\vspace*{-9 truemm}
\hspace*{-7.5 truemm}
\includegraphics[height=13cm,width=10cm,angle=0]{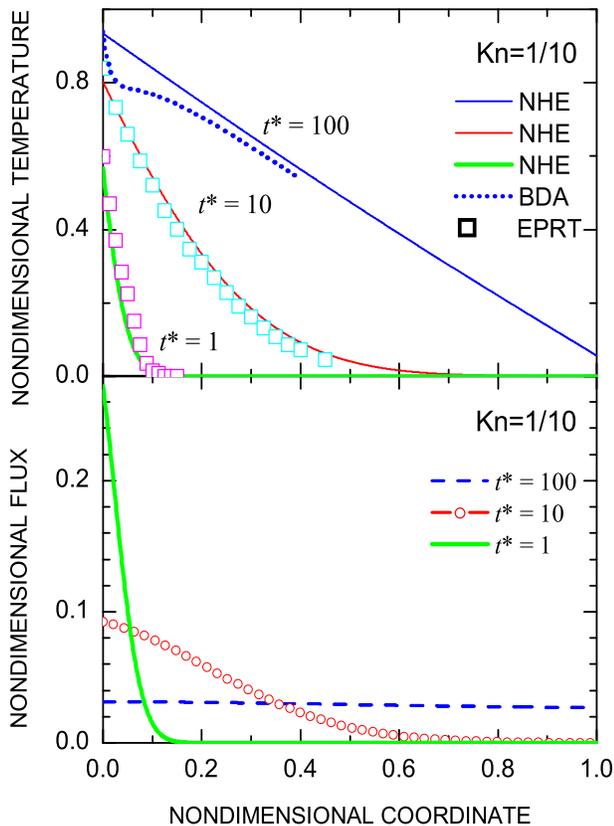}
\vspace*{-13 truemm}
\caption{\label{fig:Figure3} Profiles of $\theta(\xi,t^\ast)$ (upper panel) $\phi(\xi,t^\ast)$ (lower panel) as predicted by NHE. The open squares represent the prediction of EPRT (for $t^\ast =1 \mbox{ and }10$) reported in Ref. \cite{ChenASME:02}. The dotted curve, derived from a plot in Ref. \cite{ChenASME:02} is the output (for $t^\ast =100$) of BDA. }
\end{figure}

Let us consider now a slab for which ${\rm Kn}=0.1$, and focus attention on the temperature profiles at the instants $t^\ast=1, 10, 100$.  An examination of Chen's EPRT plot for $t^\ast=100$, which is almost linear, shows that $\xi_0$ and $\xi_1$ are, as one would expect from the above remark concerning the boundary conditions, close to $0.07$ \cite{ChenASME:02}. By choosing $\xi_0=0.07=\xi_1$, we were led to plots that agreed well, for $t^\ast=10$ and $t^\ast=100$,  with the EPRT counterparts in Chen's work; but for $t^\ast=1$, a sufficiently close fit to the EPRT values could not be obtained without changing $\xi_0$ to 0.03. To substantiate these statements, we draw the reader's attention to the upper panel of Fig.~3, where our plots are compared with Chen's BDA curve ($t^\ast=100$) or with his EPRT plots ($t^\ast=1,10$). For the sake of avoiding clutter, we have not displayed the BDA values for $\xi>0.4$, and we have also chosen not to reproduce Chen's EPRT plot (for $t^\ast=100$), which is close to the NHE plot for all values of $\xi$ and to BDA for $\xi > 0.2$.  We would like to add here that it did not seem worthwhile to improve the fit by fine-tuning the inputs for the extrapolated endpoints. Finally, we show, in the lower panel of Fig.~3, our plots for $\phi (\xi,t^\ast)$ for three particular instants ($t^\ast=1,10,100$), using the same values of $\xi_0$ as those mentioned above (0.03, 0.07 and 0.07, respectively). The plots for $t^\ast= 10$ and  $t^\ast= 100$ are in exceedingly good agreement with Chen's EPRT plots; the shape of the flux plot at $t^\ast= 1$ accords with its EPRT counterpart, but its initial amplitude is larger; we believe that this is due, at least in part, to inappropriate choices for the extrapolated endpoints.  

It seems fair to conclude, on the strength of the foregoing evidence, that NHE affords an exceptionally simple, stand-alone strategy for studying transient heat conduction in a system where precise values of the extrapolated endpoints are not needed. On the basis of the results presented in Fig.~3, it seems safe to suggest that this requirement is satisfied when ${\rm Kn}\leq 1/10$ and $t^\ast \geq 10$. For other situations, where large jumps at one or both boundaries become inevitable, Eq.~(1) can still supply the diffusive contribution to BDA.

Having illustrated the performance of an analytically tractable form of the new heat equation, we now state its general form, 
\begin{equation}
\partial_t T(x,t)=\bigl [(1-e^{-\alpha t})^2 \kappa \partial_{x}-u_0e^{-\alpha t}\bigr ]\partial_x T(x,t), 
\end{equation}
and hasten to add that the equation is new only in the context of heat conduction. If $T$ is replaced with $F$ (the probability density in coordinate space), and $\kappa$ with $D$ (the diffusion coefficient), Eq.~(5) can be identified with the equation, derived first by Ornstein and van Wijk (O\&vW) \cite{OrnsWijk:33}, for describing the time evolution of the probability density of a Brownian particle that starts with an initial velocity $u_0$; their equation reduces, if $u_0$ has a Maxwellian distribution, to an equation of the same form as Eq.~(1) \cite{SanMSanc:80}. Our reason for working with Eq.~(1) was the resulting reduction in labor. Referring to their equation, O\&vW stated that this is the ``diffusion'' equation which holds for all time intervals; and that it reduces, in the limit $t\to 0$ to $\partial_tT=-u_0e^{-\alpha t}\partial_{x}T$, describing purely ballistic behavior, and to the parabolic diffusion equation when $t\to\infty$. Following O\&vW, we point out that Eq.~(5) itself is a special case of the equation
\begin{equation}
\partial_t T(x,t)=\textstyle{1\over 2}a(t)\partial_{xx}T(x,t)-b(t)\partial_{x}T(x,t), 
\end{equation}
whose solution is Gaussian with a mean $\int_{0}^{t}b(t_1)dt_1$ and a variance $\int_{0}^{t}a(t_1)dt_1$.

It seems fit to call Eq.~(5) the {\it exact\/} ballistic-diffusive equation, and to remark that, even in the literature on Brownian motion, the analogs of Eqs.~(1) and (5) have appeared only rarely \cite{SanMSanc:80, Fox:78}. We believe, for reasons given below, that Eq.~(5) will mimic all the essential features of EPRT. 

A great merit of Majumdar's contribution \cite{Maju:93} is that it provides a concrete basis for pursuing the analogy between phonon-mediated heat conduction and particle diffusion in terms of the relevant equations. In the original formulation of EPRT, the speed of phonons is taken to be a constant, but the mean-free time is viewed as an $\omega$-dependent quantity, where $\omega$ ($0\leq \omega \leq \omega_D$) denotes the frequency. However, in the simplified version used in Refs.\cite{JoshMaju:93,ChenPRL:01,ChenASME:02} and under scrutiny here, the $\omega$-dependence of the mean free time is ignored. With a constant mean free time (denoted here by $\overline{\tau}$), EPRT can be converted, through integration  over $\omega$, into an equation for $g(x,\mu,t)\equiv \int_{0}^{\omega_D}[I_\omega (x,\mu,t)/\overline{v}] d\omega$ that has the same form as LBE with isotropic scattering, which may be written as $[\partial_t+\mu\partial_x]\psi(x,\mu,t)=\alpha({1\over 2}{\cal P}-1)\psi(x,\mu,t)$, where ${\cal P}\equiv \int_{-1}^{1}d\mu$. This implies that the internal energy $U(x,t)={\cal P}g(x,\mu,t)$ and the heat flux $q(x,t)=\overline{v}{\cal P}g(x,\mu,t)$ are the analogs of $F(x,t)={\cal P}\psi(x,\mu,t)$ and $J(x,t)\equiv \overline{v}{\cal P}\psi(x,\mu,t)$, respectively. To accommodate the concept of temperature, we recall Chen's definition \cite{ChenASME:02}, $T(x,t)=U(x,t)/C$, and stress that a different definition would invalidate the rest of the argument. Our next task is to show that $F_{\rm RT}(x,t)$ may be replaced, to all intents and purposes, by $F_{\rm BM}(x,t)$, where the suffixes RT and BM serve to specify the results pertaining to radiative transfer and Brownian motion, respectively, and it is to this task that we turn now.

Brownian motion, when modeled as an Ornstein-Uhlenbeck process \cite{UhleOrns:31,OrnsWijk:33}, differs radically from the situation portrayed by LBE \cite{Razi:94}. However, one is usually interested, not in the distribution in phase space, but in $F(x,t)$ and $J(x,t)$, and there is good reason to believe that the deviation between $F_{\rm BM}$ and $F_{\rm RT}$ is significantly smaller than that between the parent transport equations, which are really poles apart. To substantiate this claim, we examine two situations: unbounded motion, and the Milne problem, where the diffusing particles occupy the half-space $x>0$, and $x=0$ is an absorbing boundary. 

Consider a stream of particles which are released, at $x=x_i$ with the same initial velocity ${\bf u}_0$ (parallel to the $x$-axis). It has long been known that $F_{\rm BM}(x,t;x_i)$, the fundamental solution of the diffusion analog of Eq.~(5), which was found before the equation itself \cite{UhleOrns:31}, is a Gaussian, with a time-dependent mean and a time-dependent variance. For particles obeying LBE, $F_{\rm RT}(x,t;x_i)$ can be approximated, through an artifice called the cumulant approximation (CA), by a Gaussian   \cite{XuCaiLaxAlfa:02}; here we draw attention to the fact that, for isotropic scattering with no capture, the expression for $F_{\rm RT}(x,t;x_i)$ becomes, when CA is imposed, {\it identical\/} with $F_{\rm BM}(x,t;x_i)$. 
 
We move next to the Milne problem, which is exactly soluble for three linear transport equations \cite{Razi:94}: The Klein-Kramers equation, the single-relaxation time approximation to the Boltzmann equation, and LBE. Though each equation describes a different physical system, the respective plots of $F(x,t)$ against $x$ have the same overall appearance; indeed, within the lowest-order diffusion approximation, all three imply a rectilinear plot with the same extrapolated endpoint; higher-order approximations, or the exact solutions, do reveal residual differences. Given this background, one expects the temperature profile predicted by Eq.~(5) to be close, but not identical, to that found by using EPRT. The results presented above bear out this expectation. We would also like to mention the possibility that our approach might be brought still closer to EPRT by making a choice for $a(t)$ and $b(t)$ in Eq.~(6) different from that which leads to Eq.~(5). 

All that remains is to discuss the boundary conditions to be imposed on the solutions of Eqs.~(1), (5) and (6). Since there is some disagreement about this matter in the field of heat conduction  \cite{ChenASME:02}, we will content ourselves by drawing attention to a few relevant contributions concerning Milne's problem \cite{KRN:82, MenoSahn:85, MenoKumaSahn:86}; particularly relevant in this context is the finding that the extrapolated endpoint is a time-dependent quantity.  

It does not seem rash to conclude, on the basis of the results and arguments presented above, that the new heat equation will provide a very convenient basis for modeling phonon-mediated heat conduction at any spatial or temporal scale.

\end{document}